\documentclass[twocolumn,pre,showpacs,superscriptaddress]{revtex4}
\usepackage{graphicx}%
\usepackage{color} 
\usepackage{transparent}
\usepackage{psfrag}
\usepackage{dcolumn}
\usepackage{amsmath}
\usepackage{amssymb}
\usepackage{latexsym}
\usepackage{hyperref} 
\usepackage{booktabs}
\usepackage{latexsym}
\begin{document}

\def\bea{\begin{eqnarray}}
\def\eea{\end{eqnarray}}
\def\beq{\begin{equation}}
\def\eeq{\end{equation}}
\def\f{\frac}
\def\k{\kappa}
\def\e{\epsilon}
\def\ve{\varepsilon}
\def\be{\beta}
\def\D{\Delta}
\def\h{\theta}
\def\t{\tau}
\def\a{\alpha}

\def\cDa{{\cal D}[X]}
\def\cD{{\cal D}[x]}
\def\cL{{\cal L}}
\def\cLo{{\cal L}_0}
\def\cLa{{\cal L}_1}

\def\Re{{\rm Re}}
\def\sj{\sum_{j=1}^2}
\def\rk{\rho^{ (k) }}
\def\rek{\rho^{ (1) }}
\def\cek{C^{ (1) }}
\def\rz{\rho^{ (0) }}
\def\rt{\rho^{ (2) }}
\def\rtb{\bar \rho^{ (2) }}
\def\trk{\tilde\rho^{ (k) }}
\def\trek{\tilde\rho^{ (1) }}
\def\trz{\tilde\rho^{ (0) }}
\def\trt{\tilde\rho^{ (2) }}
\def\r{\rho}
\def\tD{\tilde {D}}

\def\s{\sigma}
\def\kb{k_B}
\def\bF{\bar{\cal F}}
\def\F{{\cal F}}
\def\la{\langle}
\def\ra{\rangle}
\def\nn{\nonumber}
\def\up{\uparrow}
\def\dn{\downarrow}
\def\S{\Sigma}
\def\dg{\dagger}
\def\d{\delta}
\def\p{\partial}
\def\l{\lambda}
\def\L{\Lambda}
\def\G{\Gamma}
\def\o{\Omega}
\def\w{\omega}
\def\g{\gamma}

\def\jv{ {\bf j}}
\def\jr{ {\bf j}_r}
\def\jd{ {\bf j}_d}
\def\noi{\noindent}
\def\a{\alpha}
\def\d{\delta}
\def\p{\partial} 

\def\la{\langle}
\def\ra{\rangle}
\def\e{\epsilon}
\def\n{\eta}
\def\g{\gamma}
\def\break#1{\pagebreak \vspace*{#1}}
\def\hf{\frac{1}{2}}

%\title{Active Brownian particles with velocity dependent friction: Entropy production and steady state response} 
\title{Active Brownian particles: Entropy production and fluctuation-response} 
\author{Debasish Chaudhuri}
\email{debc@iith.ac.in}
\affiliation{Indian Institute of Technology Hyderabad,
Yeddumailaram 502205, Andhra Pradesh, India
}

\date{\today}

\begin{abstract}
Within the Rayleigh-Helmholtz model of active Brownian particles activity is due to a non-linear velocity dependent force. In the presence of external trapping potential or constant force, the 
steady state  of the system breaks detailed balance producing a net entropy.  Using  molecular dynamics simulations, we obtain the  probability distributions of entropy production in these steady states. The distribution functions obey fluctuation theorems for entropy production. Using the simulation, we further show that the steady state response function obeys a modified fluctuation-dissipation relation. 
\end{abstract}
\pacs{05.40.-a, 05.40.Jc, 05.70.-a}
%\keywords{Fluctuation phenomena statistical physics, Brownian motion, Entropy thermodynamics}

\maketitle

\section{Introduction}
Active systems perform out of equilibrium dynamics by generating motion utilizing energy from their environment. 
This is unlike non-equilibrium state of passive particles, where the system is driven by external forces.
Examples of active system range from moving animals, to motile 
cells, motor proteins, and artificial active Brownian particles (ABP)~\cite{Romanczuk2012,Vicsek2012},
e.g.,  self-propelled colloids~\cite{Howse2007,Zheng2013}, nano-rotors~\cite{Nourhani2013}, 
vibrated granular particles~\cite{Feitosa2004,Joubaud2012}.
Generation of self-propulsion is often 
expressible in terms of non-linear velocity dependent forces that lead to non-zero mean speed at steady state~\cite{Romanczuk2012}.

Properties of small systems, in or out of equilibrium, are describable within the framework of stochastic 
thermodynamics~\cite{Sekimoto1998,Bustamante2005,Seifert2012}. Probability distributions of work done, or entropy production are shown to obey fluctuation 
theorems in driven passive systems, e.g., of small assembly of nano-particles, colloids, granular matter, and
polymers~\cite{Jarzynski2011,Jarzynski1997,Crooks1999,Wang2002,Liphardt2002,Feitosa2004,Narayan2004,Kurchan2007}.
While the mean entropy production in such processes remain positive, occasional fluctuation of negative entropy production is not ruled out~\cite{Evans1993,Gallavotti1995,Lebowitz1999}. 
The stochastic entropy production by particles is associated with their trajectories~\cite{Seifert2005,Seifert2008}. 
Fluctuation theorems have been verified in experiments on colloids~\cite{Wang2002, Blickle2006, Speck2007},
granular matter~\cite{Joubaud2012}, and used to find out the free energy landscape of 
 RNA~\cite{Liphardt2002,Collin2005}. 
Fluctuation theorems have also been derived for models of molecular motors~\cite{Seifert2011,Lacoste2011,Lacoste2009}, and used to determine
autonomous  force or torque generation by them~\cite{Hayashi2010,Hayashi2012}. 
Recently, fluctuation theorems for entropy production have been extended for ABPs with
velocity dependent self-propulsion forces~\cite{Ganguly2013}.
On the other hand, the non-equilibrium steady states (NESS) of driven passive Brownian particles are characterized by
response functions that obey modified fluctuation-dissipation relations (MFDR) in terms of steady state correlations~\cite{Cugliandolo1994,Speck2006,Baiesi2009,Prost2009,Seifert2010,Verley2011,Chaudhuri2012}.
Theoretical predictions in this context were verified experimentally~\cite{Blickle2007,Gomez-Solano2009}. 

 In this paper, we consider the Rayleigh-Helmholtz model~\cite{Romanczuk2012} of active Brownian particles (ABP) 
 where activity is generated via a non-linear velocity dependent force.
 Starting from underdamped Langevin equations, 
 we derive fluctuation theorems for entropy production by ABPs.  We perform molecular dynamics simulations in the presence of Langevin thermostat to  obtain
 probability distributions of entropy production to find good agreement with the detailed fluctuation theorem. 
 Finally we characterize non-equilibrium steady states of ABPs in terms of a modified fluctuation-dissipation relation. % (MFDR).   
 
%%%%%%%%%%%%%%%%%%%%%%%%%%%%%%%%%%%%%%%%%%%%%%%%%%%%%%%%%%%%%%%%%%%%%%%%
\section{Model}
The dynamics of an ABP in the presence of a velocity dependent active force $F(v)$ can be described in terms of the Langevin equations of motion
\bea
\dot x &=& v \nn\\
\dot v &=&  -\g v + \eta(t) + F(v) - \p_x U(x) + f(t). 
\label{lange}
\eea 
The Langevin heat bath is characterized by the viscous dissipation $-\g v$ and  Gaussian white noise $\eta(t)$ obeying $
\la \eta(t) \ra =0$, $\la \eta(t) \eta (t')\ra = 2  D_0 \d(t-t')$ with $D_0=\g \kb T$.  
Here $T$ denotes an effective temperature representing both thermal, and non-thermal fluctuations that may arise from
chemical processes leading to activity. 
In the above equation $U(x)$ denotes a conservative potential, and $f(t)$ a time-dependent control force.  
We use particle mass $m=1$ throughout this paper. %$F(v)$ is a velocity dependent force that can be used to model activity. 

The generation of activity by $F(v)$ can be seen easily considering $U(x)=0=f(t)$.  
In the over-damped limit, the mean velocity is obtainable from the relation $\g \la v \ra - F(\la v \ra)=0$.  
Within the Rayleigh-Helmholtz model $F(v) = a v - b v^3$ with $a>\g$. This leads to three possible fixed-points  for the 
steady state mean velocity  $\la v \ra = 0, \pm v_0$ with $v_0 = \sqrt{(a-\g)/b}$, among which $\la v \ra=0$
is unstable and $\pm v_0$ are stable fixed points. At small velocities, $v<v_0$, velocity dependent force 
$g(v)=F(v)-\g v = b(v_0^2-v^2)v$ pumps energy into the kinetic degrees of freedom to generate self propulsion~\cite{Romanczuk2012}.
This model of ABPs has been successfully used to analyze the bidirectional motion of microtubule interacting with NK11 motor-proteins
that generate active drive hydrolyzing the chemical fuel ATP~\cite{Badoual2002, Endow2000}.

The  Fokker-Planck equation corresponding to Eq.(\ref{lange}) is given by 
\bea
\p_t P(x,v,t) &=& -\p_x (v P) - \p_v[ g(v)  + \bF  ] P \crcr 
&& + D_0 \p_v^2 P \equiv - \nabla. \jv
\label{fp_eq}
\eea
where $ \nabla = (\p_x, \p_v)$ and $\bF = f(t) - \p_x U$. For a time-independent external force $f$, one may express the total current
$\jv = \jr + \jd$ with $\jr = (v P, \bF P)$ the time-reversible part of the
phase-space probability current, $\jd =  (0, g(v) P - D_0 \p_v P)$ the dissipative part of the current.
The detailed balance condition, obeying microscopic time-reversal symmetry, is satisfied if $\jd = (0,0)$ and $\nabla.\jr = 0$~\cite{Risken1989,Sarracino2013}.
The breakdown of time-reversal symmetry leads to entropy production. Thus we consider the 
detailed balance condition, and its break down in the following. 

%%%%%%%%%%%%%%%%%%%%%%%%%%%%%
\subsection{Equilibrium detailed balance}
\label{detailed}
%%%%%%%%%%%%%%%%%%%%%%%%%%%%%

The condition $\jd = (0,0)$ implies 
\bea
\p_v P(x,v) = \f{g(v)}{D_0} P(x,v) 
\label{jd0}
\eea
which has a solution 
\bea
P(x,v) = p(x) \exp[ - \phi(v)/D_0] 
\label{pxv1}
\eea
where
$\phi(v)$ is a velocity dependent potential such that $g(v) = -\p_v \phi(v)$. 
The other condition $\nabla.\jr = 0$ can be written as,
\bea
v \p_x P(x,v) + \bF \p_v P(x,v)=0
\label{jr0}
\eea
in which using $P(x,v) = p(x) \exp[ - \phi(v)/D_0]$ one obtains a solution
\bea
%p(x) = p_0 e^{[U(x)-fx]} e^{g(v)/v D_0}.
p(x) = p_0 \exp\left[-\f{g(v)}{v D_0} \, \int \bF  dx \right].
\label{px1}
\eea
If the force $\bF$ is conservative, $\bF = -\p_x U$, the solution has a normalizable form $p(x)=p_0 \exp(U(x)\,g(v)/vD_0)$.
For passive particles $g(v)=-\g v$ leads to Boltzmann distribution $p(x) = p_0 \exp(-U(x)/\kb T)$.

On the other hand if $\bF$ contains a non-conservative force $f$ the solution $p(x)$ is proportional to $\exp(-fx\, g(v)/vD_0)$, which 
is not normalizable as $\int_{-\infty}^\infty dx   \exp(-fx \, g(v)/vD_0)$ is not bounded above. Thus non-conservative force does not support
a detailed balance steady state. The requirement that conservative force, not the non-conservative one, supports microscopic reversibility is
shown in Ref.~\cite{Tome2010}, considering a many particle system.

As we show now, even conservative force, $\bF = -\p_x U$, does not allow detailed balance in ABPs.
Using the solution given by Eq.s (\ref{pxv1}) and (\ref{px1}) in Eq.(\ref{jd0}) one gets
a condition 
\bea
g(v) = - \p_v \phi(v) + \p_v \left( \f{g(v)}{v} \right) U(x).
\label{db}
\eea
Since, $g(v)=-\p_v \phi(v)$, the above condition is satisfied only if  $g(v) \propto v$, or $U(x)=0$. 
For passive Brownian particles, $g(v)=-\g v$ and conservative force always leads to equilibrium detailed balance. 
Due to non-linear velocity dependence in $g(v)$,  for ABPs in potential trap Eq.(\ref{db}) is not satisfied, 
and thus detailed balance is not obeyed. 

To summarize the discussion in this section, microscopic reversibility for ABPs may be broken either by imposing
non-conservative external force $f$, or by trapping the ABPs in conservative external potential $U(x)$. Both these
conditions, therefore, would lead to entropy production in ABPs, and are considered in this paper.

Within the Rayleigh-Helmholtz model $g(v) = (a-\g) v - b v^3$, and detailed balance is obtained if both $f=0$ and $U=0$, i.e., $\bF=0$. 
Eq.(\ref{jr0}) implies $\p_x P(x,v)=0$, which is 
automatically satisfied by the solution (\ref{pxv1}) with $p(x)=$\,constant. Thus one gets a equilibrium-like 
solution for the Rayleigh-Helmholtz model 
\bea
P_s(v) = {\cal N} \exp[-\phi(v)/D_0]
\eea 
where ${\cal N}$ is the normalization constant, and $\phi(v)=\psi(v)+\g v^2/2$ with $\psi(v)=-(a/2) v^2 + (b/4) v^4$ a velocity-dependent
double- well potential characterizing the self propulsion force $F(v)=-\p_v \psi(v)$ of the Rayleigh-Helmholtz model.
The minima of the potential $\phi(v)$ are at $\pm v_0$.
%$\phi(v) = -(bv^2/2)(v_0^2-v^2/2)$. 
%This solution remains valid in the absence of external force and potential.

\subsection{Non-equilibrium steady states}
\label{sec:ness}
The non-equilibrium steady state in the presence of a constant external force $f$, and absence of potential $U=0$, may be solved easily by noting that
the force may be incorporated by redefining the velocity-dependent potential to $\phi(v) - f v$. The corresponding steady
state distribution is
\bea
P_s(v) =  {\cal N} \exp[-\{\phi(v) - fv\}/D_0].
\label{eq:ps_vf}
\eea 
A part of the total entropy change between two steady states is the difference in stochastic system entropy 
$s = -\kb \ln P_s$~\cite{Crooks1999, Seifert2005}, as will be discussed in the next section, and thus calculation of steady state
distributions is important in the context of transient fluctuation theorems.

\begin{figure}[t]%[htbp]
\begin{center}
\input{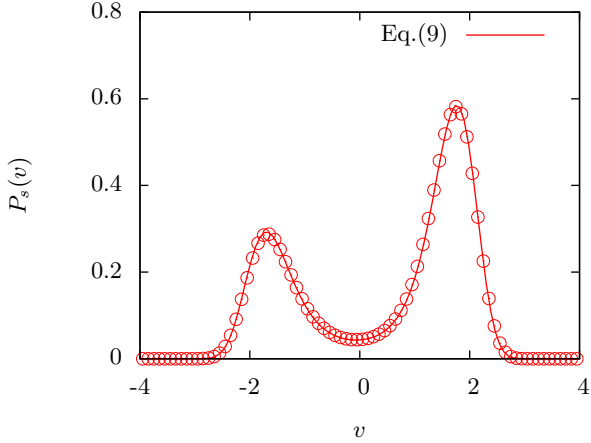}
\caption{(Color online) 
Steady state probability distribution $P_s(v)$ for ABPs %with $F(v)=a v-b v^3$ where $a=4$, $b=1$, 
under a constant external force $f=0.2$. %, and used $D_0=1$. 
%We used $\kb T=1$, $D_0=1$.
Points are from MD simulations, and the line is a plot of Eq.(\ref{eq:ps_vf}).}
\label{fig:psv}
\end{center}
\end{figure}

The Rayleigh-Helmholtz ABPs may also be brought into non-equilibrium steady state by trapping them within a conservative 
potential $U(x)$.  The analytic form of the corresponding steady state solution for general $U(x)$ is not known. 
Thus we use numerical simulations to calculate these distributions.

We perform molecular dynamics (MD) simulations using the standard
velocity-Verlet algorithm with a time-step $\d t = 0.01 \t$, where $\t=1/\g$, and keep the temperature constant at $T=1.0 (D_0/\g\kb)$
via a Langevin thermostat. %We treat the velocity dependent self propulsion force $F(v)$ in the same footing as other forces. 
The simulation method  for ABPs is validated by calculating the steady state velocity distribution under constant external force and
comparing it against Eq.(\ref{eq:ps_vf}) (see Fig~\ref{fig:psv}). In all our simulations we used $F(v)=a v - b v^3$ with $a=4$ and $b=1$.
Also, unless otherwise specified, we used the noise strength $D_0=1$. % in all our simulations.

\section{Entropy production}
The Langevin equation of the Rayleigh-Helmholtz model of ABPs, obeys energy conservation. 
Multiplying Eq.(\ref{lange}) by velocity $v$ and integrating over a small time interval ${\t_0}$ one obtains~\cite{Sekimoto1998} %the first law~\cite{Sekimoto1998}
\bea
\D E =  \D W + \D q, 
\label{1st}
\eea 
where $\D E$ denotes the change in mechanical energy  $ E= (1/2) v^2 + U(x)$, 
$\D W = \int^{\t_0} dt\, v. f(t)$ the work done on the ABPs by external force $f(t)$, 
and $\D q = \D Q + \D Q_m$ the total energy absorbed by the mechanical degrees of freedom of the ABPs:
(a) from the Langevin heat bath $ \D Q= \int^{\t_0} dt\, v. (-\g v + \eta) $, and  (b) from the self-propulsion
mechanism $\D Q_m = \int^{\t_0} dt\, v. F(v)$.  

In a system of conventional passive Brownian particles, the stochastic entropy production in any process has two components. One is the entropy change in the 
system $\D s$ where the stochastic system-entropy is expressed as $s = -\kb \ln P_s$ with $P_s$ denoting steady state distribution. The 
other contribution comes from the change in entropy in the heat-bath, $\D s_r = -\D q/T$~\cite{Seifert2005}. A direct extension of this idea
to ABPs would mean $\D s_r = -\D q/T$ with $\D q = \D Q + \D Q_m$. However, as we show below, $\D s_r $ for ABPs has further
extra contributions coming from the mechanism of active force generation and its coupling to the mechanical forces~\cite{Ganguly2013}. 

Consider the time evolution of an ABP from $t=0$ to $\t_0$ through a path defined by $X=\{x(t), v(t), f(t) \}$. 
The motion on this trajectory involves interaction of the particle with Langevin heat bath, and the presence of 
self propulsion force $F(v)$. Microscopic reversibility means the probability of such a trajectory is the same
as the probability of the corresponding time-reversed trajectory. Entropy production requires break down of such microscopic reversibility.

Let us first consider the transition probability 
$p_i^+(x',v',t+\d t | x,v,t)$ for an infinitesimal section of the trajectory evolved 
during a time interval $\d t$, assuming that the whole trajectory is made up of $i=1,\dots,N$ such segments such that $N \d t = {\t_0}$. 
The Gaussian random noise at $i$-th instant is described by $P(\eta_i) = (\d t/4\pi D_0)^{1/2} \exp(-\d t \, \eta_i^2/4 D_0)$.
The transition probability is given by 
$p_i^+ =  J_{\eta_i, v_i} \la \d(\dot x_i - v_i) \d(\dot v_i - {\cal F}_i ) \ra  
= J_{\eta_i, v_i} \int d \eta_i P(\eta_i)  \d(\dot x_i - v_i) \d(\dot v_i - {\cal F}_i )$, 
where the total force acting on the particle at $i$-th instant of time is ${\cal F}_i =  \eta_i + g(v_i) -\p_{x_i} U(x_i) + f_i$, with
$g(v_i) = F(v_i) - \g v_i$, and $J_{\eta_i, v_i} = (1/\d t)[1-\d t\, \p_{v_i} g(v_i)/2]$  (see Appendix-\ref{trajectory}). 
%Here $\dot x_i$ and $\dot v_i$ denote instantaneous velocity and acceleration respectively. 
Thus we have $p_i^+ = J_{\eta_i, v_i} (\d t/4\pi D_0)^{1/2}  \d(\dot x_i - v_i) \exp[-\f{\d t}{4 D_0} \{ \dot v_i - g(v_i) + \p_{x_i} U(x_i) - f_i \}^2] $.
%Utilizing the Markovian nature of the noise
The probability of full trajectory is ${\cal P}_+  = \prod_{i=1}^N p_i^+$.

Reversing the velocities gives us the time reversed path  $X^\dagger=\{x'(t'), v'(t'), f'(t') \} =\{x({\t_0}-t),-v({\t_0}-t), f({\t_0}-t)\}$, 
the probability of which can be expressed as ${\cal P}_-  = \prod_{i=1}^N p_i^-$ where 
$p_i^- = J_{\eta_i, v_i} (\d t/4\pi D_0)^{1/2} \d(\dot x_i - v_i) \exp[-\f{\d t}{4 D_0} \{ \dot v_i + g(v_i) + \p_{x_i} U(x_i) - f_i \}^2] $, since the velocity dependent forces are odd function of 
velocity $g(-v_i) = - g(v_i)$, and $J_{\eta_i, v_i}$ remains the same. 

The ratio of probabilities of the forward and reverse trajectories is %, thus, can be written as  
\bea
\f{{\cal P}_+}{{\cal P}_- } &=& \exp\left[{\frac{\d t}{D_0} \sum_{i=1}^N  \left( \dot v_i  + \p_{x_i} U -f_i \right)} g(v_i) \right]  \nn \\
&=& \exp\left[{\frac{1}{D_0} \int_0^{\t_0}  dt  \left( \dot v  + \f{\p U}{\p x} -f(t) \right)} g(v) \right].  \nn %\nn\\
\eea
 After simplifications the ratio can be expressed as~\cite{Ganguly2013} % (see Appendix-B of ~\cite{Ganguly2013}) 
\bea
\f{{\cal P}_+}{{\cal P}_- } 
 &=& \exp\left[ - \be \left(\D q + \D Q_{em} +\f{1}{\g} \D \psi \right) \right]
 \label{p+p-}
\eea
where $\be = 1/\kb T = \g/D_0$.  
In the above relation  $\D q = \D Q + \D Q_m$  is the heat absorbed, as identified in the context of energy conservation.
The term $\D Q_{em} = (1/\g)\int_0^{\t_0} dt\,  F(v).(f(t)- \p_x U)$ is a coupling  between 
the self-propulsion and external forces.
$\D \psi$ is the change in a self-propulsion potential defined through $F(v) = -\p_v \psi(v)$. 
%For Rayleigh-Helmholtz ABPs $\psi(v) = -(a/2) v^2 + (b/4) v^4$.
%Note the difference between $\psi(v)$ and the previously introduced
%function $\phi(v)= (\g /2) v^2 + \psi(v)$. 

%%%%%%%%%%%%%    FIGURES  %%%%%%%%%%%%%%%

\begin{figure}[t]%[htbp]
\begin{center}
\input{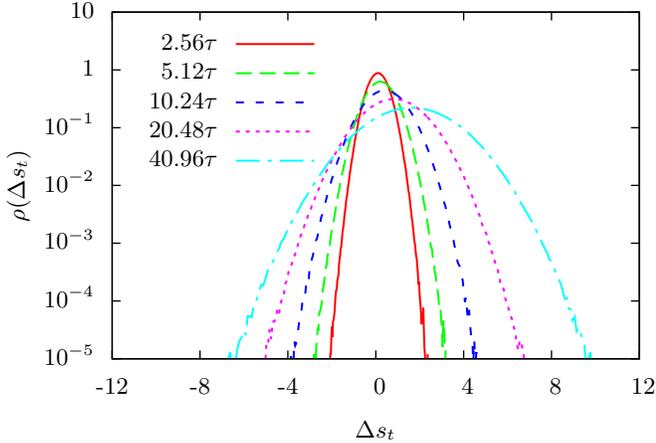}
\caption{(Color online) Probability distribution of total entropy production $\r(\D s_t)$ calculated in the presence of 
an external force $f=0.2$. %We used $\kb T=1$, $D_0=1$. %, and $F(v)=a v-b v^3$ where $a=4$, $b=1$.
The calculations are performed after collecting data over $\t_0=2.56, 5.12, 10.24, 20.48, 40.96 \,\t$. }
\label{fig:PS}
\end{center}
\end{figure}
%%%%%%%%%%%%%%%%%

The probability ratio of the forward and reverse trajectories accounts for the entropy change in 
the reservoirs ${\cal P}_+/{\cal P}_- = \exp(\D s_r/\kb)$~\cite{Seifert2005,Ganguly2013}.
Thus we have 
\bea
\D s_r = - \f{1}{T} \left(\D q + \D Q_{em} +\f{1}{\g} \D \psi \right). 
\eea  
Evidently the reservoir entropy change $\D s_r$ has contributions from two extra terms, $\D Q_{em}$ and $ \D \psi$, with respect to 
the expression $\D s_r = -\D q/T$, inferred from the behavior of passive Brownian particles. 

It is interesting to note that the active force has three contributions to entropy production. 
Origin of $\D Q_m$ in $\D q=\D Q + \D Q_m$ is direct, this is due to work done by the active force.
The contribution through energy transfer  $\D Q_{em}$ is due to coupling of velocity- dependent active force 
to mechanical forces.   
Apart from that, the mechanism of active force generation through the velocity dependent potential $\psi(v)$ also
contributes to entropy. 
The origin and meaning of these terms have easy interpretation within a simple model of active particle dynamics 
$\dot v = -\g (v-v_0)+\eta(t) + f(t)$ considered in Ref.s~\cite{Romanczuk2011,Schienbein1993}.
In this model, friction $\g$ pumps in energy if $v<v_0$, and dissipates otherwise.
The self propulsion force $F=\g v_0$ leads to $\D Q_{em} = \int dt f v_0$, 
and $\D \psi/\g = - \D (v v_0)$. 
Thus, in this case  $\D Q_{em}$ and  $\D \psi/\g$ are equivalent to work done, and change in internal energy 
for driven passive Brownian particles, respectively.

Assuming the initial and final steady state distributions as $P^i_s$ and $P^f_s$ respectively, the system entropy change is $\D s = s_f - s_i = \kb \ln(P_s^i/P_s^f)$.
Thus the total entropy production is
\bea
\D s_t &=& \D s -\f{1}{T}\left(\D q + \D Q_{em} +\f{1}{\g} \D \psi \right) \nn\\
&=&  \D s -\f{1}{T}\left(\D E -\D W + \D Q_{em} +\f{1}{\g} \D \psi \right),
\label{s_tot}
\eea
where in the last step we used the relation of energy conservation Eq.(\ref{1st}).

%%%%%%%%%%%%%%%%
\begin{figure}[t]%[htbp]
\begin{center}
\input{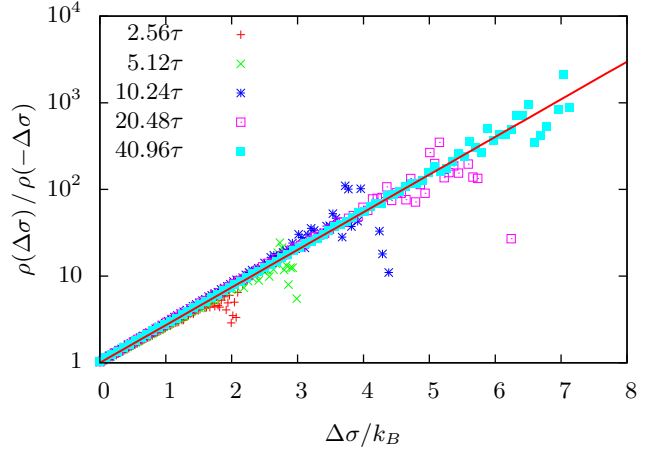}
\caption{(Color online) Ratio of probability distributions of positive and negative entropy production $\r(\D s_t = \D\s)/\r(\D s_t = -\D\s)$ calculated from 
the data in Fig.~\ref{fig:PS}.
The solid line is a plot of the function $\exp(\D \s/\kb)$.}
\label{fig:PSratio}
\end{center}
\end{figure}
%%%%%%%%%%%%%%%%%  

The probability distribution of the forward process is ${\cal P}_f = P_s^i {\cal P}_+$, and that of the reverse process
is ${\cal P}_r = P_s^f {\cal P}_-$. Thus 
\bea
{\cal P}_r / {\cal P}_f = \exp(-\D s_t/\kb),
\label{eq:ratio}
\eea
and 
$\la \exp(-\D s_t/\kb) \ra = \int {\cal D}[X] {\cal P}_f  \exp(-\D s_t/\kb) = \int {\cal D}[X] {\cal P}_f \, ({\cal P}_r / {\cal P}_f) =1$.
This relation is known as the integral fluctuation theorem~\cite{Kurchan2007} 
and implies a positive  entropy production on an average $\la \D s_t \ra  \geq 0$,  .

Eq.(\ref{eq:ratio}) can be used to obtain the detailed fluctuation theorem for the probability distribution of entropy production $\r(\D s_t)$~\cite{Crooks1999,Ganguly2013},
\bea
 \f{\r(\D \s)}{\r(- \D \s)} = e^{\D \s/\kb},
\eea 
where $\D\s$ denotes an amount of total entropy $\D s_t$ produced over a time interval $\t_0$.
In the following, using MD simulations we calculate the steady state probability distributions of 
total entropy productions $\r(\D s_t)$ and hence test the detailed fluctuation theorem.

\subsection{Detailed balance state}
In the absence of external potential $U(x)=0$, and force $f(t)=0$, the system obeys detailed balance as has been shown in Sec.~\ref{detailed}.
Let us denote the initial and final points on a trajectory evolved over a time ${\t_0}$ by $(x_i,v_i)$ to $(x_f,v_f)$. 
In this case, the heat absorbed $\D q = \D E = (v_f^2 - v_i^2)/2$, and the steady state distribution $P_s = {\cal N} \exp[-\phi(v)/D_0]$ where
$\phi(v) = (\g/2) v^2 +\psi(v) $ with $\psi(v)=  - (a/2) v^2 + (b/4) v^4$. The corresponding  entropy change in the system is 
$\D s /\kb = \D \phi/D_0 = \D \psi/D_0 + (\be/2) (v_f^2 - v_i^2) $ with $\be = \g/D_0$. Thus the total entropy change is
\bea
\f{\D s_t}{ \kb} &=& \f{\D s}{\kb} - \be \left(\D q + \f{1}{\g}\D \psi \right) \nn\\
&=& \f{\D \phi}{D_0}  - \f{\be}{2} (v_f^2 - v_i^2) - \f{\D \psi}{D_0} \nn\\
&=& 0,
\eea
as expected due to detailed balance. There is no difference between the initial and final steady states, and
the probabilities of forward and reverse trajectories are the same. %, and hence $\D s_t=0$.

%%%%%%%%%%%%%%%%
\begin{figure}[t]%[htbp]
\begin{center}
\input{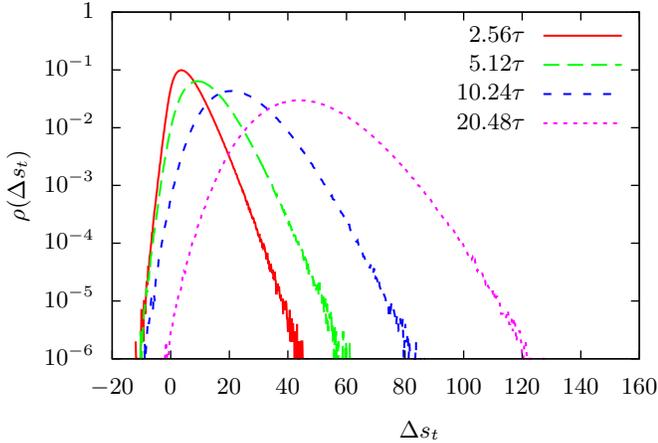}
\caption{(Color online) Probability distributions of total entropy productions $\r(\D s_t)$ 
calculated in the presence of an external harmonic potential trap $U(x)=(1/2) \w_0^2 x^2$ with $\w_0^2=5$.
The calculations are performed after collecting data over $\t_0=2.56, 5.12, 10.24, 20.48 \,\t$. } %, 40.96
\label{fig:PS_U}
\end{center}
\end{figure}
%%%%%%%%%%%%%%%%%  

%%%%%%%%%%%%%%%%
\begin{figure}[t]%[htbp]
\begin{center}
\input{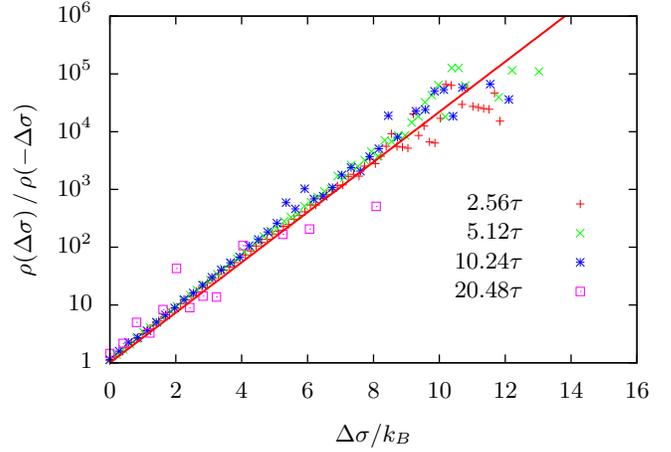}
\caption{(Color online) Ratio of probability distributions of  positive and negative entropy productions $\r(\D s_t = \D \s)/\r(\D s_t = -\D \s)$ 
calculated from the data shown in Fig.~\ref{fig:PS_U}.
The solid line shows a  plot of $\exp(\D \s/\kb)$.}
\label{fig:PSratio_U}
\end{center}
\end{figure}
%%%%%%%%%%%%%%%%%  

\subsection{NESS with constant force}
The simplest non-equilibrium steady state producing entropy is attained in the presence of a constant 
external force, breaking the detailed balance condition for ABPs. % of $U=fx$. 
In this case $f \neq 0$ and external potential $U(x)=0$. 
We assume a trajectory from $(x_i,v_i)$ to $(x_f,v_f)$ 
evolves over time ${\t_0}$. 
The heat absorbed is $\D q = \D E - \D W = (v_f^2 - v_i^2)/2 - f (x_f - x_i)$.
The steady state distribution  is given by (Eq.(\ref{eq:ps_vf}))
$P_s = {\cal N} \exp[-\{\phi(v) - f v\}/D_0]$ where  $\phi(v) = (\g/2) v^2 +\psi(v) $ with $\psi(v)=  - (a/2) v^2 + (b/4) v^4$. Thus the 
system entropy change $\D s /\kb = (\D \phi -f \D v)/D_0 = \D \psi/D_0 + (\be/2) (v_f^2 - v_i^2) - (f/D_0)(v_f-v_i)$. The total entropy
change is 
\bea
\f{\D s_t}{ \kb} &=& \f{\D s}{\kb} - \be \left(\D q + \D Q_{em} + \f{1}{\g}\D \psi \right) \nn\\
&=& -\f{f}{D_0}\left [(v_f-v_i) +\int^{\t_0} dt F(v)  \right]+\be f (x_f - x_i)  \nn\\
\label{eq:fst}
\eea
where in the last step we used the identity $\be \D Q_{em} = (f/D_0) \int^{\t_0} dt F(v)$. 

In Fig.~\ref{fig:PS} we show the probability distributions of entropy production $\r(\D s_t)$ 
calculated from MD simulations of Rayleigh-Helmholtz ABPs at $f=0.2$, 
using Eq.(\ref{eq:fst}) for the expression of $\D s_t$. 
The distributions are calculated after collecting data over
various time periods ${\t_0}$. Appreciable probability of negative entropy production is clearly visible. 
With increase in ${\t_0}$, the distributions broaden and 
the peak positions shift towards higher values of entropy. 
From each curve, one can extract the 
ratio of probabilities $\r(\D \s)/\r(- \D\s)$ with $\r(\D \s) = \r(\D s_t = \D \s)$ and $\r(-\D \s) = \r(\D s_t = -\D \s)$.
As is shown in Fig.~\ref{fig:PSratio}, this ratio obeys the detailed fluctuation theorem $\r(\D \s)/\r(- \D\s) = \exp(\D \s/\kb)$.

\subsection{ABPs in potential trap}

A system of Rayleigh-Helmholtz ABPs if trapped by an external potential $U(x)$ (keeping $f=0$) gets into a NESS. 
This is unlike passive Brownian particles that still remains at equilibrium with probability distribution described 
in terms of  Boltzmann weight $\exp[-\be U(x)]$. 
As we have seen in Sec.~\ref{sec:ness}, the steady state probability density $P_s(x,v)$ in this case 
is not analytically obtainable for a general $U(x)$ and noise strength $D_0$. We perform MD simulations to find $P_s(x,v)$. 
For a trajectory between $(x_i,v_i)$ and $(x_f,v_f)$ evolved over a time ${\t_0}$, the corresponding 
change in the system entropy is thus calculated using the numerically obtained probability distributions,
and the relation $\D s = \kb \ln [P_s(x_i,v_i)/P_s(x_f,v_f) ]$. 
The change in the reservoir entropy is given by
\bea
\f{\D s_r}{\kb} = -\be\left[ \D E - \f{1}{\g} \int^{\t_0} dt F(v) \p_x U(x) + \f{\D \psi}{\g}  \right],
\eea 
where $E = v^2/2 + U(x)$, and as before, for any function $\chi(x,v)$ the change $\D \chi(x,v) = \chi(x_f, v_f) - \chi(x_i, v_i)$.
In MD simulations, we use $U(x) = (1/2)\w_0^2 x^2$, a harmonic potential well with strength $\w_0^2=5$. % with $\w_0=1$.
Probability distribution of entropy production $\r(\D s_t)$ is shown in Fig.~\ref{fig:PS_U}.
The distribution widens, and the peak rapidly moves towards
very large values of total entropy as the measurement time $\t_0$ is increased. %However, 
The detailed fluctuation theorem is obeyed
as is shown in Fig.~\ref{fig:PSratio_U}.

%%%%%%%%%%%%%%%%%%%%%%%%%%%%%%%%%%%%%%%%%%%%%%%%%%%%%%%%%%%%%%%%%%%%%%%
\section{Linear response at NESS: modified fluctuation dissipation relation}
%%%%%%%%%%%%%%%%%%%%%%%%%%%%%%%%%%%%%%%%%%%%%%%%%%%%%%%%%%%%%%%%%%%%%%%
The steady state of the ABPs may be characterized by linear response functions.
The Fokker-Planck equation (\ref{fp_eq}) can be written as
\bea
\p_t P(x,v,t) = \cL(x,v,h) P(x,v,t) =(\cLo + f(t) \cL_1) P
\eea
where
\bea
\cLo P &=& - \p_x(v P) - \p_v \left[ g(v) -\p_x U \right] P + D_0 \p_v^2 P \nn\\
\cL_1 P &=& -\p_v P. \nn
\eea 
As it has been shown earlier, the linear response to $f(t)$ in a system at steady state described by $P_s(x,v)$ such that 
$\cLo P_s =0$ can be expressed as~\cite{Chaudhuri2012, Seifert2010, Verley2011, Agarwal1972}
\bea
\f{\d \la A(t)\ra}{\d f(t')} = \la A(t) M(t') \ra_s
\label{eq:mfdr1}
\eea 
where $\la \dots \ra_s$ indicates a steady state average, and  $M = -({1}/{P_s}) \p_v P_s$. This is a version
of modified fluctuation dissipation relation (MFDR). 

For free ABPs $U(x)=0=f(t)$, the system goes into a  detailed balance steady state
described by the distribution $P_s(v) = {\cal N} \exp[-\phi(v)/D_0]$ where $\phi(v) = -(a-\g) v^2/2 + b v^4/4$.
In this case, $M=\p_v [-\ln P_s] = g(v)/D_0=  [ - (a-\g) v + b v^3)/D_0$, and the
response function $R_A(t,t')={\d \la A(t)\ra}/{\d f(t')}$ around a steady state, where time translation
invariance is obeyed, is given by
\bea
R_A(t)
= - \f{a-\g}{D_0} \la A(t) v(0) \ra_s + \f{b}{D_0} \la A(t) v^3(0) \ra_s.
\label{mfdr1}
\eea
For the ABPs, $a > \g $ gives rise to active force generation leading to a negative coefficient of 
$\la A(t) v(0) \ra_s$ in the MFDR. Given that the fluctuation dissipation theorem for passive 
Brownian particles is $R_A(t) = \be \la A(t) v(0) \ra_{eq}$,  within equilibrium the temperature can be
expressed as the ratio $\kb T = \la A(t) v(0) \ra_{eq} / R_A(t)$. 
For ABPs, even in a detailed balance state,
the effective temperature $T$ is not expressible as a simple ratio of fluctuation $\la A(t) v(0) \ra_s$ and response $R_A(t)$,
and the coefficient of $\la A(t) v(0) \ra_s $ can not be interpreted as an effective negative temperature.

\begin{figure}[t]%[htbp]
\begin{center}
\input{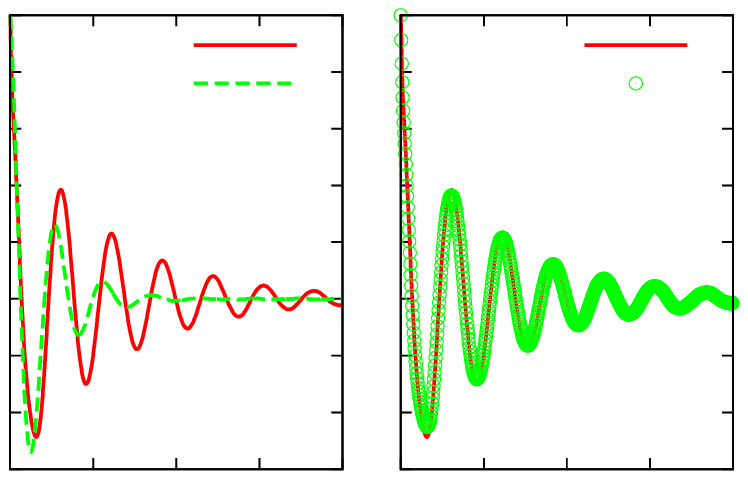}
\caption{(Color online) Response functions and steady state fluctuations. ($a$)~Direct MD evaluation of response function
for Rayleigh-Helmholtz ABPs, and passive Brownian particles within a harmonic trap of strength $\w_0^2=5$.
($b$)~ Comparison of response function of ABPs $R_v(t)$ against steady state fluctuations as given by 
the right hand side of Eq.(\ref{eq:mfdr}).}
\label{fig:Rv}
\end{center}
\end{figure}
%%%%%%%%%%%%%%%%%

In order to use the expression Eq.(\ref{eq:mfdr1}), one requires the detailed knowledge of the steady state
probability distribution. Interpreting the Gaussian noise $\eta(t)$ in the same footing as the externally applied forces, and by 
expressing the observable $A(x(t),v(t))$  as a functional $A[\eta(t)]$ of the noise history, the 
response function can also be written as~\cite{Speck2006} 
\bea
R_A(t-t') = \left\la\f{ \d A[\eta]}{\d \eta(t') } \right\ra = \f{1}{2D_0} \la A(t) \eta(t') \ra . %\nn\\
\eea 
Using the Langevin equation to replace $\eta(t')$, for ABPs under a potential $U(x)$ one finds 
\bea
R_A(t) &=& \f{1}{2D_0}  [   \la A(t) \dot v(0) \ra  -   \la A(t) g[v(0)] \ra  \nn \\
&& + \la A(t) \p_x U[x(0)] \ra    ].
\eea
Let us now focus our attention on velocity response $R_v(t)$ in NESS.
Utilizing causality and time translation symmetry at the NESS the above expression can be written as~\cite{Sarracino2013}
\bea
R_v(t) &=& -\f{1}{2 D_0} [  \la g[v(t)]  v(0) \ra  +   \la v(t) g[v(0)] \ra\nn\\
&& - \la \p_x U[x(t)] v(0) \ra -\la v(t) \p_x U[x(0)] \ra ]
\eea
where $g[v(t)] = -\g v(t) + F[v(t)]$.
For harmonic traps $U(x)=(1/2)\w_0^2 x^2$, the above expression further simplifies, as $\la x(t) v(0) \ra = - \la v(t) x(0) \ra$, to 
\bea
R_v(t) = -\f{1}{2D_0}  [\,  \la g[v(t)]  v(0) \ra  +   \la v(t) g[v(0)] \ra \, ].
\label{eq:mfdr}
\eea
Even for $U=0$ this relation holds, but the system goes to 
a detailed balance state, in which, due to time reversal symmetry $\la g[v(t)]  v(0) \ra = \la v(t) g[v(0)] \ra$, and thus %one obtains
\bea
R_v(t) = -\f{1}{D_0}  \la v(t) g[v(0)]  \ra  
\eea 
which is the same as Eq.(\ref{mfdr1}) for velocity response.
For passive free particles, $g(v) = -\g v$, and one gets back the equilibrium response function $R_v(t) = \be \la v(t) v(0) \ra = \exp(-t)$.
However, when placed within a harmonic trap they are expected to show an oscillatory response. 

Note that $\d \la v(t) \ra = \int_{-\infty}^t R_v(t-t') \d f(t') d t'$
and replacement of the perturbing force $\d f(t') $ by a Diract-delta function $\d(t')$ gives $\d \la v(t) \ra = R_v(t)$. 
Thus in MD simulations, velocity response is calculated by following the change in velocity due an impulsive force of unit magnitude.
In Fig.~\ref{fig:Rv}($a$) we show the comparison between the response functions $R_v(t)$ evaluated from 
MD simulations of harmonically trapped passive Brownian particles with that of the  Rayleigh-Helmholtz ABPs. Activity clearly leads to longer lasting 
oscillations. In the non-equilibrium  steady state that the ABPs maintain, 
our simulations show $\la g[v(t)]  v(0) \ra \neq \la v(t) g[v(0)] \ra $ which is due to the absence of time-reversal symmetry. 
We find a good agreement between the directly calculated response function $R_v(t)$ with that of the steady state fluctuations expressed
by the right hand side of Eq.(\ref{eq:mfdr}) (see Fig.~\ref{fig:Rv}($b$)). The correlation functions are calculated from a separate MD simulation performed in the 
absence of external force.

\section{conclusion}
Using molecular dynamics simulations, 
we obtained probability distributions of entropy production in non-equilibrium steady states of the Rayleigh-Helmholtz ABPs.
We identified the conditions under which ABPs break detailed balance and start to produce entropy. 
We showed that the entropy production obeys the detailed fluctuation theorem. Further, we verified a modified fluctuation-dissipation relation for the
steady state response. Given the close relation of the Rayleigh-Helmholtz model to the bidirectional motion of microtubules influenced by NK11 motors~\cite{Badoual2002},   
our predictions are amenable to experimental verification.

\acknowledgements
D.C. thanks Abhishek Dhar and Sriram Ramaswamy for useful discussions, and Arnab Saha for a critical reading of the manuscript.

\appendix
\section{Probability of a trajectory}
\label{trajectory}
It is simpler to consider an over-damped Langevin dynamics first. Let us assume the position of a particle evolves via
\bea
\g \dot x = \eta(t) + \F
\eea
where $\F$ is the total non-stochastic force acting on the particle, and the Gaussian white noise is characterized by
$\la \eta(t) \ra =0$, $\la \eta(t)\eta(0) \ra = 2 D_0 \d(t)$ with $D_0= \g \kb T$.
 Discretizing the equation with $t=i\, \d t$, using Stratonovich rule, 
\bea
x_{i}=x_{i-1} + \f{\be D}{2} (\F_i + \F_{i-1}  )\d t + \xi_i \d t
\label{apt}
\eea
where $D=\kb T/\g$ and $\xi_i = \eta_i/\g$. 
The Gaussian random noise $\xi(t)$ follows the distribution $P(\xi_i) = (\d t/4 \pi D) \exp(-\d t \xi_i^2/4 D)$. Thus the transition
probability $P(x_i | x_{i-1}) = J_{\xi_i,x_i}\, P(\xi) $ where the Jacobian 
\bea
J_{\xi_i,x_i} = {\rm det} \left( \f{ \p \xi_i}{\p x_i }\right) = \f{1}{\d t}\left( 1- \f{\d t}{2 \g} \p_{x_i} \F_i\right).
\eea 
Using Eq.(\ref{apt}) to replace $\xi_i$, we find
\bea
P(x_i | x_{i-1}) = J_{\xi_i,x_i}\, \sqrt{\f{\d t}{4 \pi D}} e^{-\f{\d t}{4 D} \left[ \f{x'-x}{\d t} + \be D \F \right]^2 }.
 \eea
 This transition probability is easily obtainable from the probability of velocity  calculated at $i$-th instant
 $\la \d (\dot x -v) \ra$ where  $v = (\eta +\F)/\g$ ,
 \bea
 \la \d (\dot x -v) \ra &=&  \int d \xi  \sqrt{\f{\d t}{4 \pi D}} e^{-\f{\d t}{4 D}  \xi^2} \d (\dot x -v) \nn\\
 &=& \sqrt{\f{\d t}{4 \pi D}} e^{-\f{\d t}{4 D} \left[ \dot x + \be D \F \right]^2 }.
 \eea 
 Identifying $\dot x =( x_i - x_{i-1})/\d t$, the transition probability, or the probability of a segment of the
 trajectory between $(x_{i-1},t)$ and $(x_i,t+\d t)$ is $P(x_i | x_{i-1}) = J_{\xi_i,x_i}\, \la \d (\dot x -v) \ra$. 
 The whole trajectory is obtainable by adding a series of such segments. 
 The probability weight associated with the whole trajectory is ${\cal P}^+ = \prod_i P(x_i | x_{i-1})$~\cite{Seifert2008}. 
 
 A direct extension of this idea to under-damped Langevin equation is straightforward. The dynamics is described by
 \bea
 \dot x &=& v \nn\\
 \dot v &=& g (v) + \eta(t) + \F 
 \eea
 where $g(v)$ contains all the velocity-dependent forces, and $\F$ denotes the velocity-independent forces.
 Similarly as in the above calculation, the probability of $i$-th segment of the trajectory 
 $p^+_i \equiv P(x_i, v_i | x_{i-1},v_{i-1}) = J_{\eta_i,v_i} \,\la \d (\dot x -v) \d (\dot v - \{g(v) +\F \} ) \ra $ which gives
 \bea
 p^+_i = J_{\eta_i,v_i}  \,\d (\dot x -v) \sqrt{\f{\d t}{4 \pi D_0}} e^{-\f{\d t}{4 D_0} \left[ \dot v +\g v -  \F \right]^2 },
 \eea
 where~\cite{Ganguly2013}
 \bea
 J_{\eta_i,v_i} = \f{1}{\d t} \left( 1 -\f{\d t}{2} \p_{v_i} g(v_i)\right).
 \eea
The probability associated with a full trajectory is ${\cal P}^+ = \prod_i p^+_i$.

\bibliographystyle{prsty}

\end{document}